\documentclass[lettersize,journal]{IEEEtran}
\usepackage{amsmath,amsfonts}
\usepackage{algorithmic}
\usepackage{algorithm}
\usepackage{array}
\usepackage[caption=false,font=normalsize,labelfont=sf,textfont=sf]{subfig}
\usepackage{textcomp}
\usepackage{stfloats}
\usepackage{url}
\usepackage{verbatim}
\usepackage{graphicx}
\usepackage{cite}
\begin{document}

\title{An Extended Nonlinear Stability Assessment Methodology For Type-4 Wind Turbines via Time Reversal Trajectory}


\author{Sujay~Ghosh, Mohammad Kazem Bakhshizadeh,  Guangya Yang and~Łukasz Kocewiak
\thanks{Manuscript received xx xx. This work was supported in part by DTU strategic alliance stipend funding.}
\thanks{S. Ghosh (corresponding author) is with Ørsted Wind Power A/s, Nesa Allé 1, 2820, Denmark (e-mail:  sujgh@orsted.com).}
\thanks{M. K. Bakhshizadeh, is with Ørsted Wind Power A/s, Nesa Allé 1, 2820, Denmark (e-mail: modow@orsted.com).}
\thanks{G. Yang is with the Technical University of Denmark, Anker Engelunds Vej 1, 2800, Denmark (e-mail: gyyan@dtu.dk).}
\thanks{Ł. Kocewiak is with Ørsted Wind Power A/s, Nesa Allé 1, 2820, Denmark (e-mail: lukko@orsted.com).}}



\maketitle

\begin{abstract}
As the integration of renewable energy generation increases and as conventional generation is phased out, there is a gradual decline in the grid's strength and resilience at the connection point of wind turbines (WTs). Previous studies have shown that traditional grid-following controlled converters exhibit deteriorating dynamic characteristics and may result in an unstable system when connected to a weak grid. Due to the limitations of linear analysis, transient stability investigations are necessary. However, existing methods, such as standalone time-domain simulations or analytical Lyapunov stability criteria, have drawbacks, including computational intensity or excessive conservatism. Our prior research proposed an innovative approach to estimate the system boundary - a time-limited region of attraction (TLRoA), using a hybrid linearised Lyapunov function-based method and the time-reversal technique to compensate for the known limitations. However, in that work, the accuracy of the estimated TLRoA was not investigated, i.e. the TLRoA was not compared against a forward simulated region of attraction, and the sensitivity of the system parameters on the TLRoA was not explored. Moreover, the framework did not consider nonlinear control elements such as PLL saturation. In this paper, we not only build upon our previous work and propose directions that address these gaps but also enhance its effectiveness by introducing optimal sampling to improve further the speed of estimating the TLRoA. Furthermore, the stability boundary is verified using time-domain simulation studies in PSCAD.
\end{abstract}

\begin{IEEEkeywords}
Adaptive sampling, Region of attraction, saturation, Time trajectory reversal, Transient stability assessment, Wind turbine converter system.
\end{IEEEkeywords}

\section{Introduction}
\IEEEPARstart{T}{he} global increase in offshore wind power plants (OWPPs) installation is driven by the demand for sustainable and clean energy sources \cite{1}. However, OWPPs, being situated far from shore and having high power output relative to grid strength, present various technical challenges, including system stability concerns. Notably, research on converter-interfaced wind turbines has highlighted their tendency to exhibit rapid and intricate transient responses during grid disturbances \cite{2} \cite{3}. Specifically, in grid-following control schemes, the phase-locked loop (PLL) controls play a key role in driving this complex behaviour \cite{4}.

The stability assessment of WPP connections has traditionally involved the utilisation of linear techniques such as eigenvalue analysis \cite{5}\cite{6} or impedance-based stability analysis \cite{7} \cite{8}. These approaches assume that the entire system, encompassing the wind turbine (WT) and the interconnected power system, demonstrates a linear response when subjected to small disturbances. Consequently, stability analysis is confined to a specific operating point. In contrast, nonlinear approaches, including time-domain simulations, equal-area criteria, and Lyapunov's direct method, among others, permit the incorporation of nonlinear characteristics and offer comprehensive asymptotic stability conclusions \cite{9} \cite{10}. An established nonlinear stability metric is the region of attraction (RoA) \cite{11}, wherein the area of the RoA provides an estimate of how stable the system is. A large RoA has a large stability margin, and the primary challenge in assessing nonlinear stability (or transient stability) is to estimate this nonlinear stability boundary - RoA.

As part of our ongoing research on system modelling and nonlinear system stability assessment methods, a novel reduced-order model (ROM) that can emulate the actual WT system behaviour has been proposed in \cite{12} \cite{13}. The ROM significantly reduces the complexity of the detailed WT model by retaining only the states of interest. Considering the challenges associated with (a) conducting numerous time-consuming simulations to analyse a wide range of system conditions, (b) developing intricate nonlinear energy functions for analytical methods that often yield conservative assessments of system RoA, and (c) necessitating expertise in data-driven techniques for optimisation and machine learning methods, a new approach was introduced in \cite{14} to evaluate the transient stability of WTs. This method combines the reverse time trajectory technique [15] with linear Lyapunov functions to estimate a time-limited region of attraction (TLRoA).

\begin{figure*}[h]
    \centering
    \includegraphics[width=16.20cm]{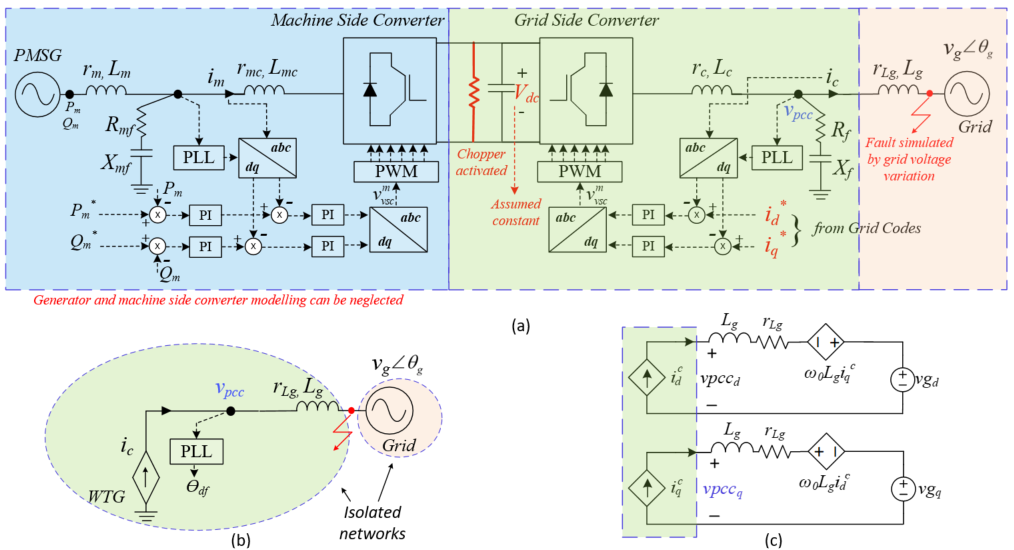}
    \caption{Wind turbine model: (a) Full topology of Type-4 wind turbine system, highlighting the actions/assumptions during faults. (b) Reduced order model (ROM) of the Type-4 wind turbine considering the actions/assumptions. Also shows the synchronisation instability of wind turbine systems during grid faults. (c) System representation of ROM in the DQ domain.}
    \label{fig6}
\end{figure*}

The TLRoA transient stability assessment methodology \cite{14} is fast and straightforward and can be quickly adopted by the industry. However, the accuracy of the estimated TLRoA was not investigated, i.e. the TLRoA was not compared against a forward simulated RoA, and the sensitivity of the system parameters on the TLRoA was not explored. Moreover, the framework did not consider nonlinear control elements such as PLL saturation. In this paper, we not only build upon our previous work and propose directions that address these gaps but also enhance its effectiveness by introducing optimal sampling to improve further the speed of estimating the TLRoA. The paper's contribution can be summarised as follows:

\begin{enumerate}
    \item Analysing the accuracy of the estimated TLRoA by comparing it against forward simulated RoA and further exploring the sensitivity of system parameters on the TLRoA.
    
    \item Improving the proposed transient stability assessment methodology by considering the neighbouring RoAs and including nonlinear control elements such as PLL saturation in the modelling framework.
    
    \item Exploring optimal sampling strategies to enhance the speed of estimating the TLRoA. This leads to more accurate and efficient RoA estimations.
\end{enumerate}

\section{Transient stability assessment}
Our previous research papers, including \cite{12} \cite{13} explored how a type-4 wind turbine can be reduced to a grid-side converter with a constant DC voltage during grid fault analysis, as shown in Fig. 1(a). It was observed that the fast inner current control dynamics could be ignored for transient stability analysis, and the shunt capacitor filter's impact could be disregarded if the current was controlled on the grid-side LCL filter. Therefore, the grid-side converter can be represented by a controlled current source with the current injection and the ramping rates obtained from relevant grid codes. Figure 1(b) shows the reduced-order representation of the WT model, and Fig. 1(c) presents the DQ-domain WT reduced-order model, for balanced grid conditions.

\subsection{System modelling and parameters}
The equivalent swing equation of the WT converter system derived in \cite{12} can be presented as,

\begin{equation}\label{SEq_1}
M_{eq} \dot{x_2} = T_{m_{eq}} - T_{e_{eq}} - D_{eq}\dot{x_1}
\end{equation}

where, $x_1= \delta$ and $x_2= \dot{\delta}$,
\begin{equation}\label{SEq_2}
\begin{aligned}
M_{eq} &= 1- k_p L_g i_d^c\\
T_{m_{eq}} 
&= k_p( \dot{\overline{r_{Lg} i_q^c}} + \ddot{\overline{L_g i_q^c}} + \dot{\overline{L_g i_d^c}} \omega_g)
+ k_i( r_{L_g} i_q^c \\
&\qquad+ \dot{\overline{L_g i_q^c}} + L_g i_d^c \omega_g) \\
T_{e_{eq}} 
&=  (k_i V_g \text{sin} x_1 + k_p \dot{V_g} \text{sin} x_1) + M \dot{\omega}_g\\
D_{eq} 
&= k_p ( V_g \text{cos}x_1 - \dot{\overline{L_g i_d^c}}) - k_i L_g i_d^c
\end{aligned}
\end{equation}

The WT system is modelled in a DQ frame rotating at a fixed frequency $\omega_0$. The equation includes several variables and constants, such as $k_p$ and $k_i$, which are the PLL controller gains, $i_d^c$ and $i_q^c$, representing the converter currents in the converter reference frame, $r_{Lg}$ and $L_g$, which denote the grid impedance, and $V_g$ and $\omega_g$, representing the grid voltage and frequency, respectively. The mathematical model for the WT system and its controls, during and after the fault, is discussed in detail in \cite{12}.

Table 1 presents the operating point of the WT converter system considered in this study.

\begin{table}[h]
\caption{SYSTEM AND CONTROL PARAMETERS \cite{16}}
\label{table}
\setlength{\tabcolsep}{3pt}
\begin{tabular}{|p{40pt}|p{105pt}|p{70pt}|}
\hline
Symbol& 
Description& 
Value \\
\hline
$S_b $& 
Rated power& 
12 MVA \\
$V_g$& 
Nominal grid voltage (L-N, pk) & 
690 $\sqrt{2/3}$ V\\
$f_g$& 
Rated frequency & 
50 Hz \\
$r_{Lg}$, $L_{g}$& 
Grid-side impedance & 
SCR=3.3, X/R=18.6 \\
$i_d^c$, $i_q^c$& 
Pre-disturbance active and reactive currents (pu) & 
1.0, 0 \\
$K_{pll}$& 
SRF PLL design: $k_p$ $k_i$ & 
0.025, 1.5 \\
$i_d^{c, ramp}$& 
Post fault active current ramp rate & 
28.4 kA/s and 42.6 kA/s\\
\hline
\multicolumn{3}{p{240pt}}{The PLL gains are chosen to obtain an oscillatory PLL behaviour.}
\end{tabular}
\label{tab1}
\end{table}

\subsection{Transient stability problem and objectives}
Following a large disturbance, the system dynamics change. 

Suppose a system has a stable post-disturbance equilibrium $x_0$ = \{$\delta(t_0)$, $\dot{\delta}(t_0)$\}, such that f($x_0$) = 0; then, as described in \cite{17}, the main problem of transient stability is whether the system can reach $x_0$ from the post-fault condition $x_f$ = \{$\delta(t_f)$, $\dot{\delta}(t_f)$\}.

The enclosed area that contains all the post-fault conditions $x_f$ from which trajectories converge to the equilibrium point $x_0$ is called the RoA, denoted by A($x_0$),

\begin{equation}
A(x_0) = {x| \lim_{t \to \infty} \Phi(t, x) = x_0 }
\end{equation}

Consequently, the objectives of transient stability assessment are to:
\begin{itemize}
\item Identify the post-disturbance condition $x_f$.
\item Evaluate if $x_f$ is within the estimated RoA of a post-disturbance stable equilibrium $x_0$.
\end{itemize}

\subsection{Estimating RoA through forward time simulations}
Typically, the system RoA has been identified through time-domain simulations, where determining the RoA requires repeating simulations over a large set of initial conditions. For example, in Fig. 2, the RoA is estimated by repeatedly simulating the system (1) with different post-fault initial conditions (no post-fault active current ramp, see Section II-E2). The green region indicates the initial conditions where the system is stable, and the blue region represents the neighbouring RoA for the adjacent equilibrium points that repeat every $\pm 2\pi$ radians. The red region corresponds to all unstable initial conditions. To estimate a smooth stability boundary, it is often necessary to simulate the system with a high resolution of initial conditions. In Fig. 2, around 3200 simulations were performed, which took approximately thirty minutes, in an Intel Core i7-CPU @ 1.10 GHz.

\begin{figure}[h]
    \centering
    \includegraphics[width=7.75cm]{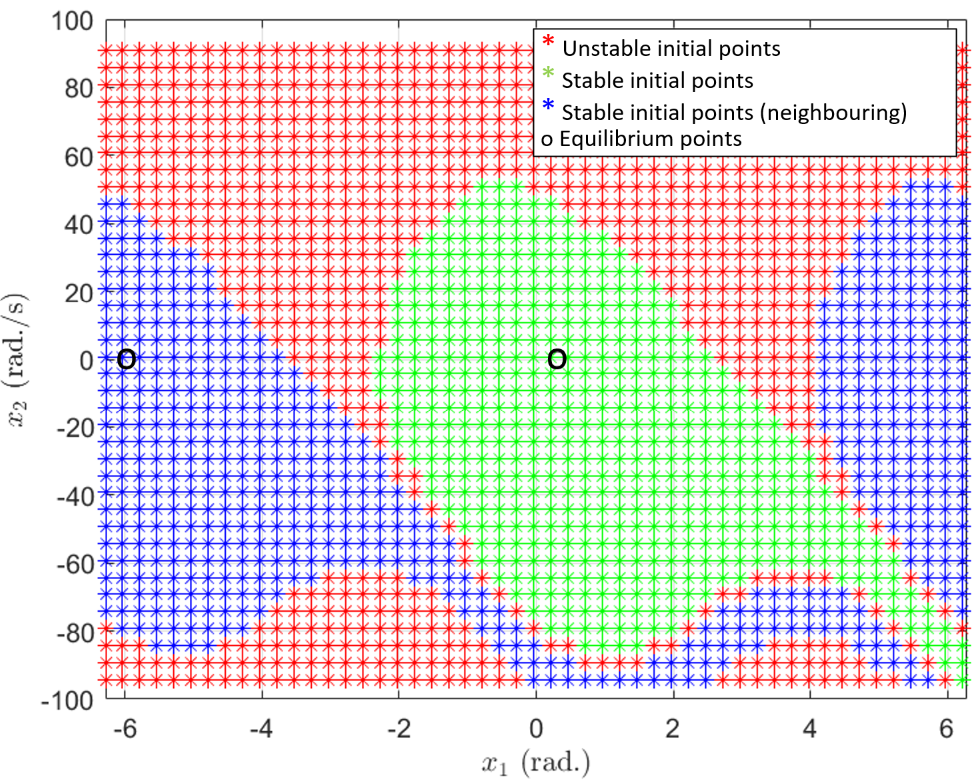}
    \caption{Forward simulated RoA, without post fault active current ramp rate.}
    \label{fig:Ramp1}
\end{figure}

Similar studies must be repeated for scenarios with different model parameters, which can be problematic for several reasons, including increased computational resource requirements, delayed results, and reduced efficiency. Hence, there is a need to explore different RoA estimation methodologies, especially considering the reduced computational burden.

\subsection{Estimating RoA through reverse time simulations}
The time-reversal technique has been extensively researched for several decades \cite{18}-\cite{21}. The application of time reversal in dynamic systems dates back to 1915 when it was initially used to analyse a three-body problem \cite{21}.
In \cite{14}, we established that as our system (1) has a unique solution, and the boundary is preserved in a homeomorphic mapping, the system can be time reversed. For example, consider a fault-ride-through scenario as presented in Fig. 3.

\begin{figure}[h]
    \centering
    \includegraphics[width=7.5cm]{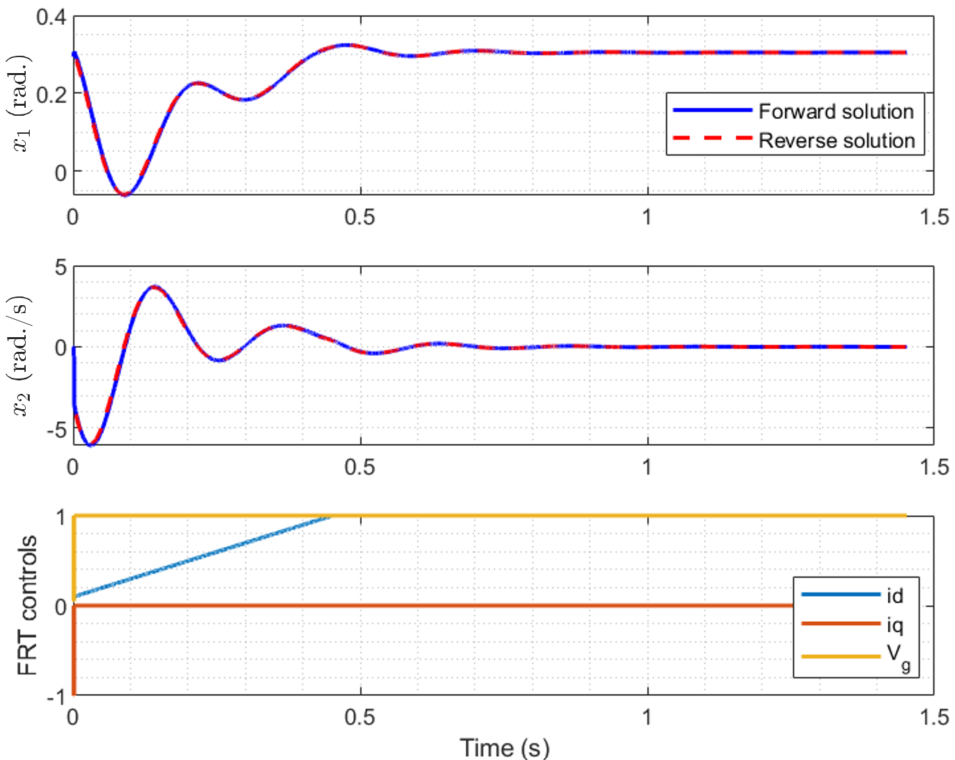}
    \caption{Introduction to Reverse time trajectory.}
    \label{fig:Ramp1}
\end{figure}

In Fig. 3, the post-fault system (1) is initially allowed to propagate in the forward direction until it reaches a tolerance band around the steady state condition. Then, for the reverse time trajectory, the final condition of the forward simulation becomes the initial condition for the reverse time simulation. The forward and reverse trajectories of system (1) are presented in Fig. 3, and it is observed that they are similar, which shows that our system (1) is time-reversible. The advantage of reverse time simulation is that, to estimate the RoA, it only needs to be performed for stable cases, which considerably reduces the number of repeated time-domain simulations.

Stability theory \cite{11} guarantees that any path within the RoA will eventually reach a state of equilibrium, but it does not provide information about the duration it takes, which could range from a few seconds to several minutes. However, in power systems, it is often necessary for the system to reach a stable state within a specific timeframe, known as the settling time. To address this requirement, the time-limited region of attraction (TLRoA) was introduced in reference \cite{14}. The TLRoA is a subset of the RoA, characterised by trajectories on its boundary that reach equilibrium at the same time, while trajectories within it reach equilibrium in less time than the specified duration.

To estimate the system TLRoA, its initial conditions must first be estimated, and this obviously must not be computed through the forward simulations. In this regard, a linearised Lyapunov function approach was proposed, which defines the tolerance band around the equilibrium point. For more information on establishing the initial conditions (initial RoA) for reverse trajectory, it is advised to refer to Section III-A in \cite{14}. Once the initial conditions are established, the system can be simulated backwards for time $t$ (in this case, 1 s) to establish the TLRoA, as shown in Fig. 4. The time $t$ is the post-fault settling time of the system, whose range can usually be obtained from the grid code requirements.

\begin{figure}[h]
    \centering
    \includegraphics[width=7.5cm]{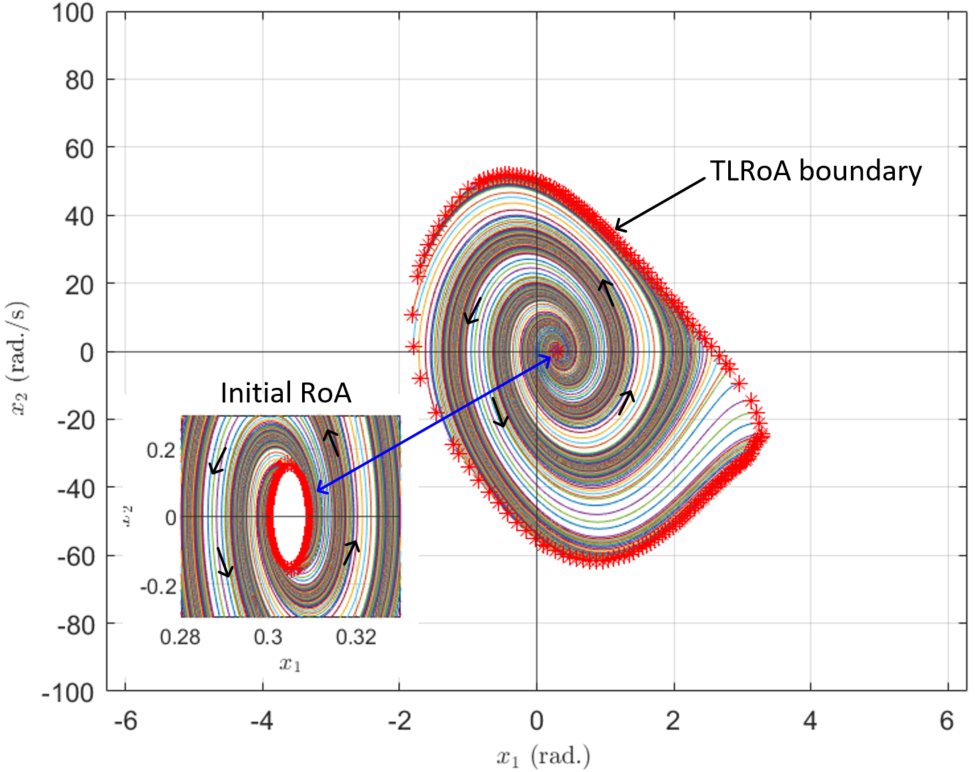}
    \caption{Estimated TLRoA of the system (1) without post-fault active current ramp, and backward simulation time of 1 sec.}
    \label{fig:Ramp1}
\end{figure}

In Fig. 4, a red asterisk on the TLRoA boundary represents a single time-domain simulation. To obtain a smooth boundary  (set of final conditions), it is necessary to take enough samples from the boundary of the initial RoA. It is essential to carefully evaluate the accuracy of the boundary estimation and consider the trade-offs between accuracy and computational resources when determining the number of simulations needed.

\subsection{Sensitivity of TLRoA in relation to system parameters}
In this section, the sensitivity of the system parameters on the TLRoA is presented and is compared against forward simulated RoA.   

\subsubsection{Backward simulation time}
As discussed earlier in Section II-D, the reverse trajectory's duration $t$ is typically determined by the settling time specified in different grid codes. In Fig. 5, we vary the duration $t$ (0.9 s, 1.0 s, and 1.1 s) and compare the resulting TLRoA with the forward simulated RoA. We observe that as the duration $t$ increases, the TLRoA estimate also grows until the trajectories hit the actual system boundary (i.e., the forward simulated RoA) and further increasing the duration no longer affects the TLRoA's expansion. Therefore, varying the backward simulation time until the TLRoA stops growing is necessary for a more accurate estimation. It must be noted that this sensitivity study is carried out just to show that the TLRoA can get closer to the forward simulated RoA; this practice is not recommended due to specific settling time requirements in the grid codes.  

\begin{figure}[h]
    \centering
    \includegraphics[width=7.5cm]{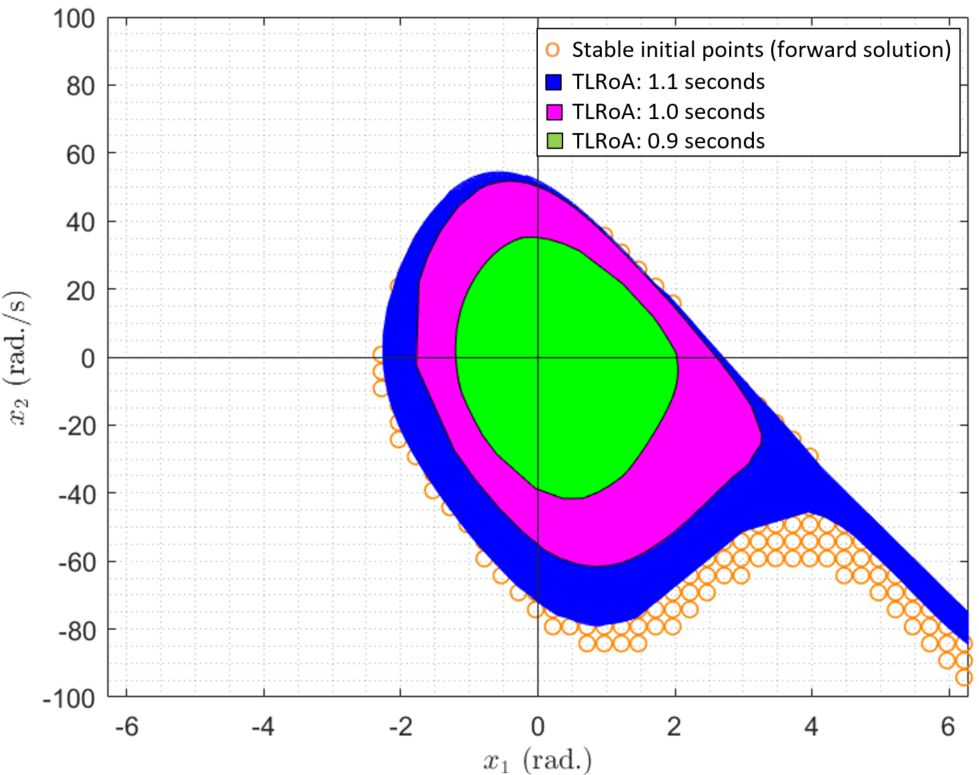}
    \caption{Estimated TLRoA of the system (1) without a post-fault active current ramp and backward simulation time of 0.9s, 1s and 1.1s.}
    \label{fig:Ramp1}
\end{figure}

\subsubsection{Post-fault active current ramp-rate}
A ramp change of active current during recovery is a common practice recommended by grid codes such as the German VDE. This ramp is typically used to avoid undue stress on wind turbines' mechanical structure, which can improve the post-fault system's stability. Previous research \cite{22} has shown that a system with a slower post-fault active current ramp rate will have a larger RoA and is, therefore, more stable than a system with faster ramp rate control. Figures 6 and 7 present the estimated TLRoA with ramp rates of 28.4 kA/s and 14.2 kA/s, respectively, and compares them to the forward simulated RoAs. The system (1) is backwards simulated for 1 s until it reaches the ramp, after which it is simulated until the ramp ends (for reference, see Fig. 3, subfigure 3).

\begin{figure}[h]
    \centering
    \includegraphics[width=7.5cm]{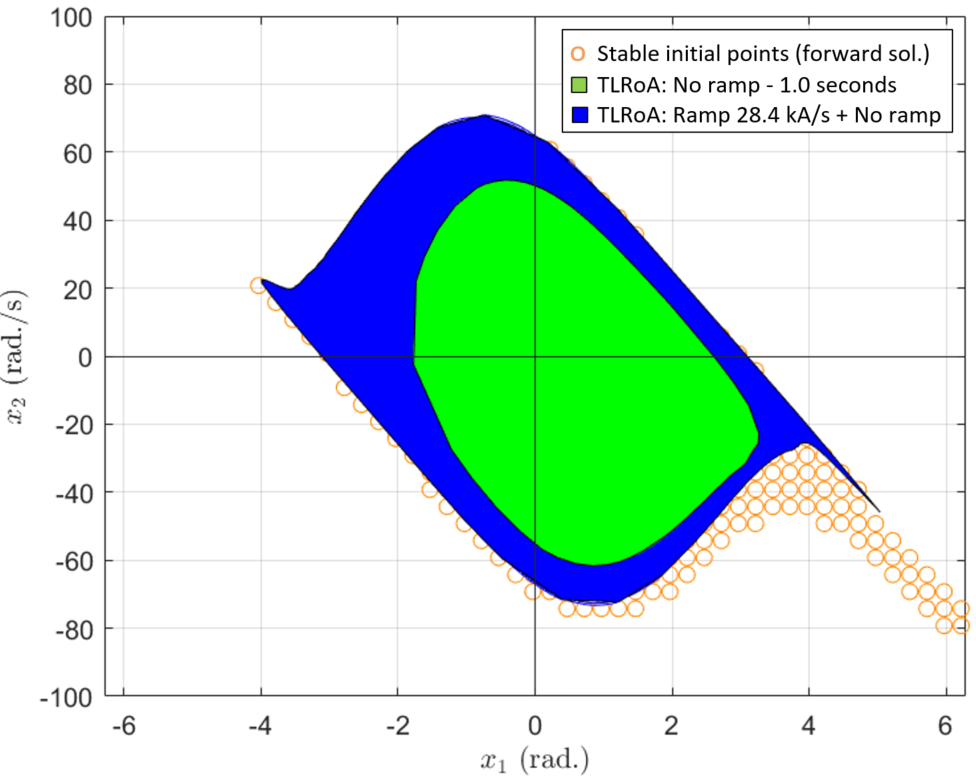}
    \caption{Estimated TLRoA of the system (1) with a post-fault active current ramp of 28.4kA/s, and backward simulation time of 1s + ramp duration.}
    \label{fig:Ramp1}
\end{figure}

\begin{figure}[h]
    \centering
    \includegraphics[width=7.5cm]{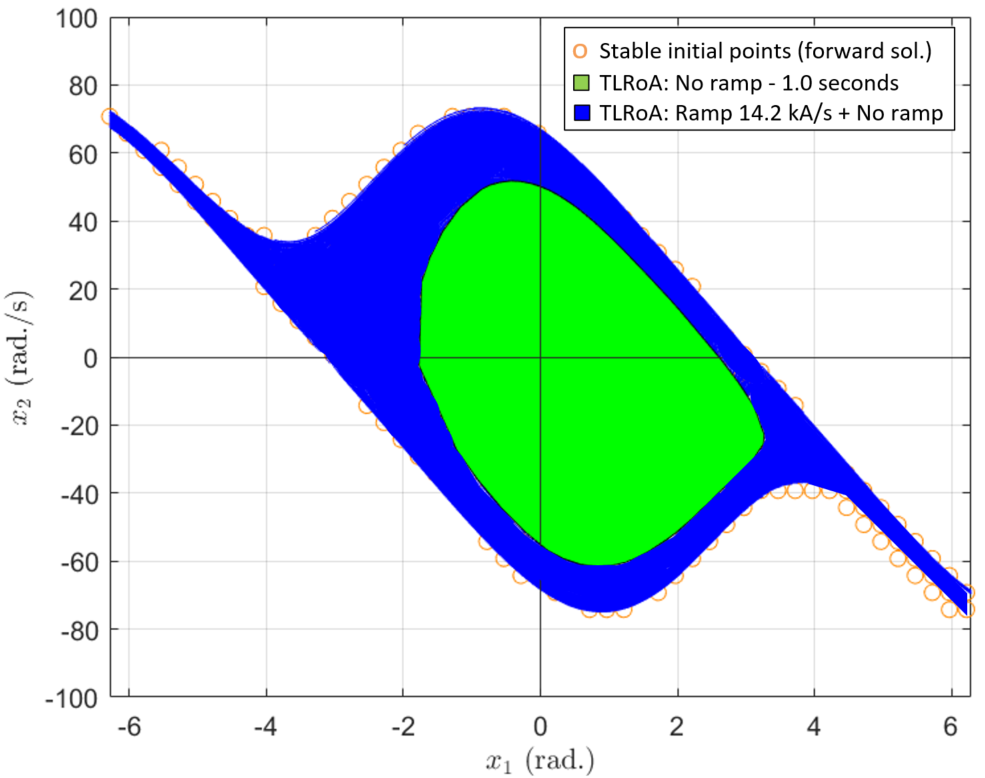}
    \caption{Estimated TLRoA of the system (1) with a post-fault active current ramp of 14.2kA/s, and backward simulation time of 1s + ramp duration.}
    \label{fig:Ramp1}
\end{figure}

In Fig. 6, a good match is obtained between the TLRoAs and the forward simulated RoAs. Also, it is observed that a slower recovery ramp rate has a larger RoA compared to a system with faster ramp rate control.

\subsubsection{Fault active current injection}
During severe grid faults, the maximum steady-state value of $i_d$ can be determined by solving $\sqrt{I_{max}^2 - i_q^2}$ = 0.45 pu, where assuming $I_{max}$ = 1.1p.u. and $i_q$ = –1.0p.u. (as specified by grid codes for a voltage less than 0.5 p.u.). However, injecting this value of $i_d$ may result in a system that is unstable or has a decreased stability margin because active power and angle are correlated, as noted in \cite{22}. To investigate this characteristic, Fig. 8 shows the estimated TLRoA with different levels of active current injection during faults. The figure demonstrates that a system with higher active current injection (i.e., $i_d$ = 0.45 pu) results in a much smaller TLRoA than a system with very low active current injection (i.e., $i_d$ = 0.01 pu).

\begin{figure}[h]
    \centering
    \includegraphics[width=7.5cm]{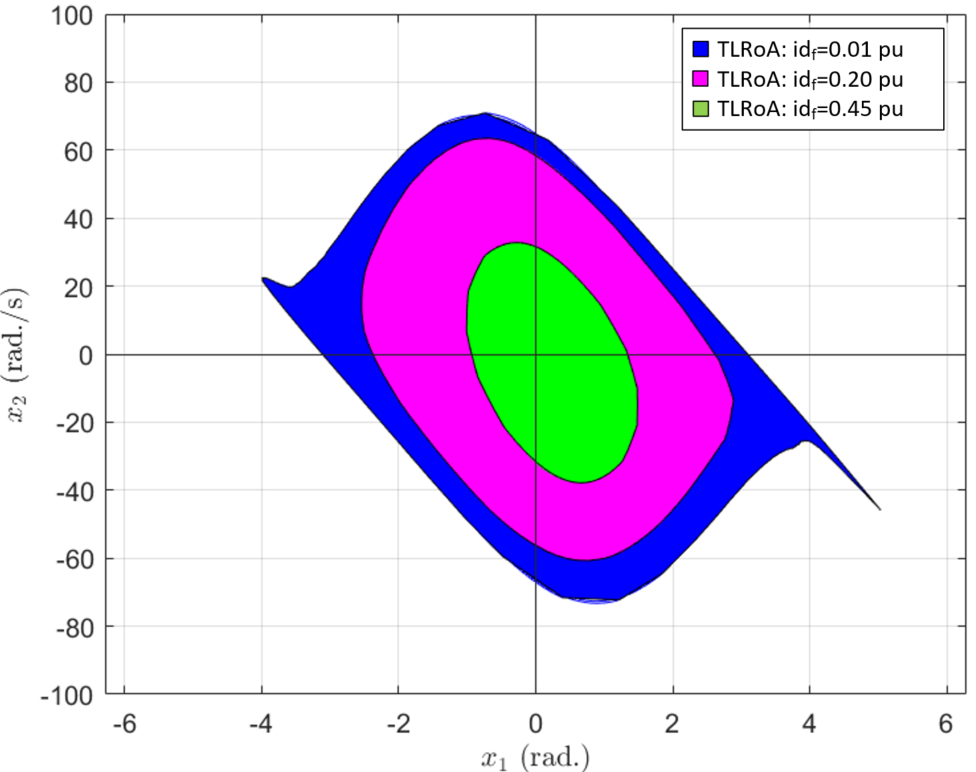}
    \caption{Estimated TLRoA of the system (1) with a post-fault active current ramp of 28.4kA/s, backward simulation time of 1s + ramp duration, and varying active current injection during fault.}
    \label{fig:Ramp1}
\end{figure}

\subsubsection{Grid short circuit ratio (SCR)}
Figure 9 presents how strong and weak grid scenarios affect the TLRoA for the system (1). The study includes two simulations, one with a grid SCR of 3.3 and another with a grid SCR of 1.1. The results indicate that the TLRoA area is much smaller in the weak grid scenario compared to the strong grid scenario. This finding is unsurprising, as a weaker grid implies less support for the system, making it less stable and more vulnerable to disturbances. The results highlight the importance of considering the grid strength when evaluating the stability of a system and provide insight into how changes in the grid parameters can impact system stability.

\begin{figure}[h]
    \centering
    \includegraphics[width=7.5cm]{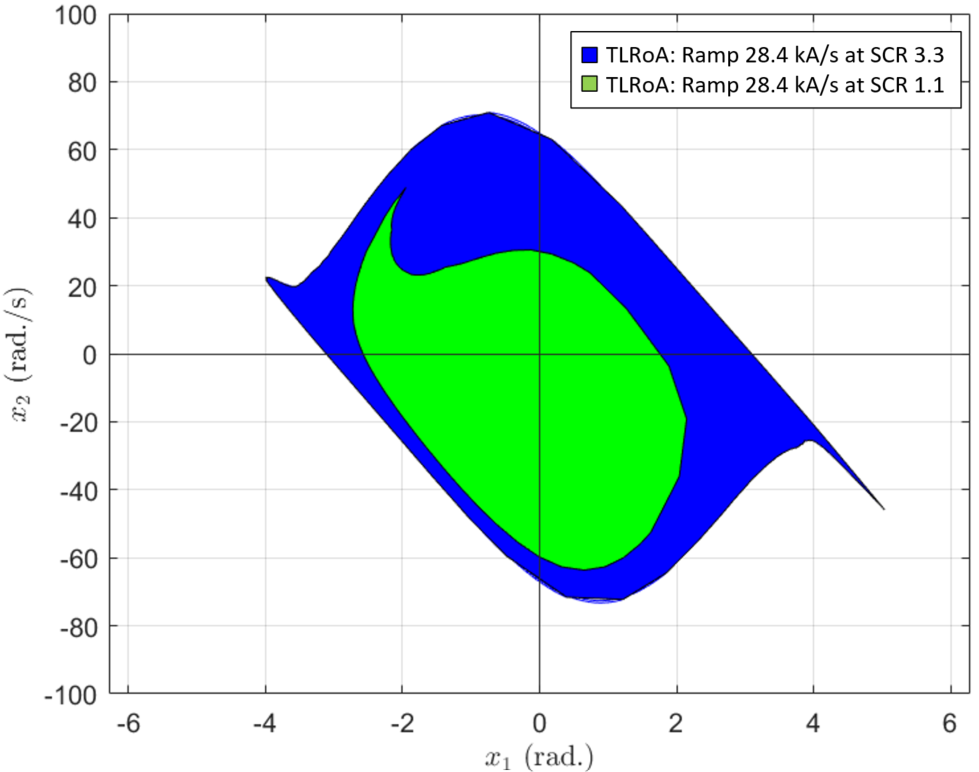}
    \caption{Estimated TLRoA of the system (1) with a post-fault active current ramp of 28.4kA/s, and backward simulation time of 1s + ramp duration, and varying grid SCR.}
    \label{fig:Ramp1}
\end{figure}

In general, the sensitivity studies revealed some crucial insights into obtaining a better estimate of the system's TLRoA. First, it was observed that varying the backward simulation time until the TLRoA stops growing results in a better estimate of the system boundary; however, the backward simulation time usually depends upon the settling time requirements. Additionally, a fresh perspective on the impact of system stability on system parameters was presented. The studies showed that a slower post-fault active current ramp rate, a lower active current injection during the fault, and a higher grid SCR result in a larger TLRoA estimate. In other words, these factors increase the stability margin of the system.

\section{Transient Stability Assessment Methodology}
This section presents the enhancements suggested to extend the approach in \cite{14} to access transient stability.    

\subsection{Need for neighbouring ROAs}
To achieve the objectives of transient stability, in our proposed method, the TLRoA represents the post-disturbance system. And a forward time-domain simulation is performed to evaluate the initial condition during the fault.

For instance, Fig. 10 presents the estimated TLRoA for system (1) with a post-fault active current ramp rate of 28.4 kA/s. Further, the forward fault trajectory is overlaid on the estimated TLRoA in delta-omega coordinates, with the PLL angle reset to $\pi$ rad. when it reaches $-\pi$ radians. In \cite{14}, it was suggested that clearing the fault at any point along the fault trajectory (red line) within the TLRoA ensures that the system will be attracted to its post-fault equilibrium point. However, if the fault is cleared outside the TLRoA, it was proposed that the system becomes unstable.

\begin{figure}[h]
    \centering
    \includegraphics[width=7.5cm]{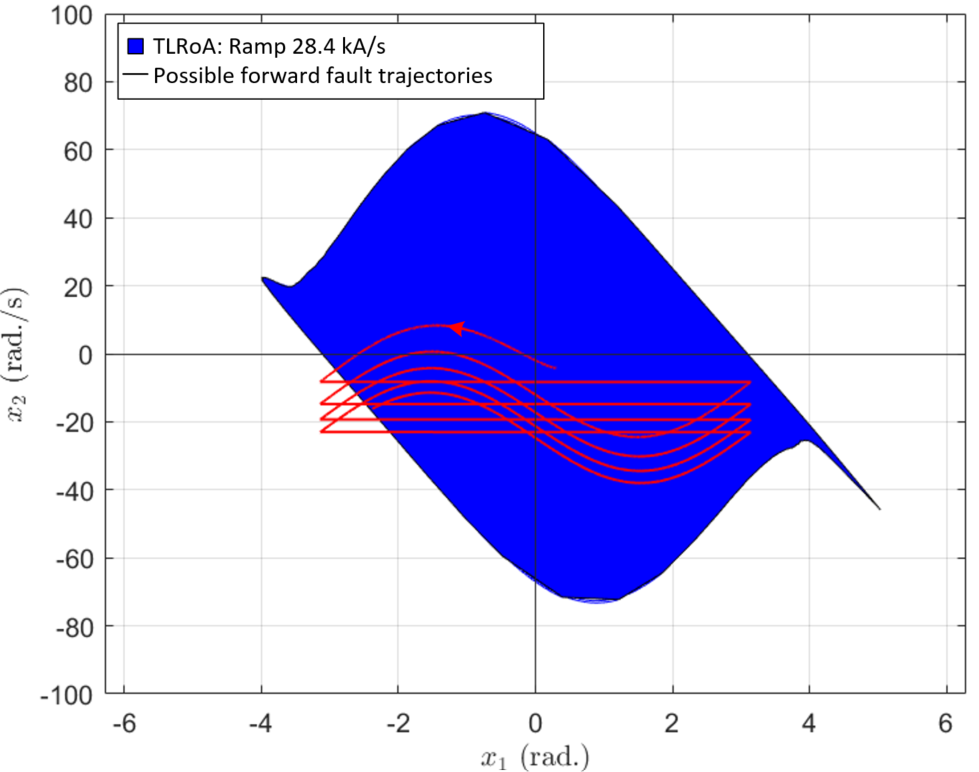}
    \caption{Transient stability assessment method as per \cite{14}.}
    \label{fig:Ramp1}
\end{figure}

The previous statement is valid only under certain circumstances. To illustrate this, let us consider a system with a very large RoA, as shown in Fig. 11, where the SCR is 3.3. The neighbouring RoAs for the adjacent equilibrium points, which repeat every $\pm 2\pi$ radians, are also included for analysis purposes. It can be observed that although the fault is cleared outside the green TLRoA, the trajectory enters the neighbouring RoA, and thus the system remains stable. However, consider a system with a small RoA, as shown in Fig. 12, where the SCR is 1.1. Here, it can be observed that the fault is cleared outside the green TLRoA; however, here, the trajectory does not enter the neighbouring RoA. Therefore, the system becomes unstable only in cases where the fault is cleared outside the TLRoA and also outside the neighbouring TLRoA. Consequently, it is essential that when analysing the transient stability, the neighbouring RoAs must be considered in the analysis.

\begin{figure}[h]
    \centering
    \includegraphics[width=7.5cm]{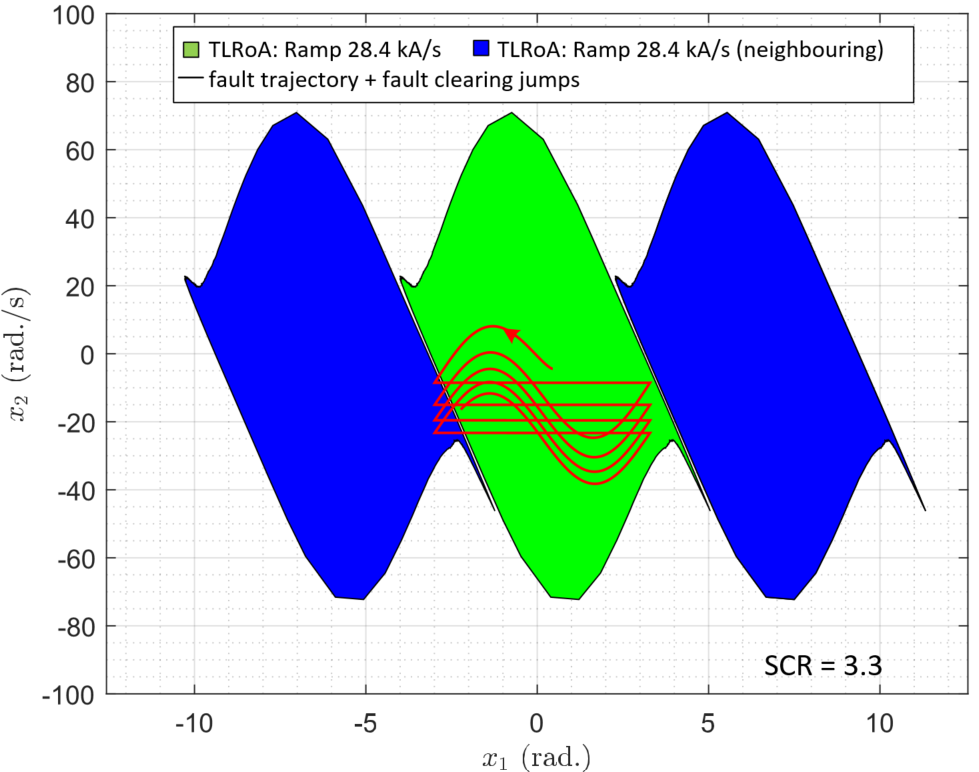}
    \caption{Transient stability assessment: highly stable system with grid SCR of 3.3.}
    \label{fig:Ramp1}
\end{figure}

\begin{figure}[h]
    \centering
    \includegraphics[width=7.5cm]{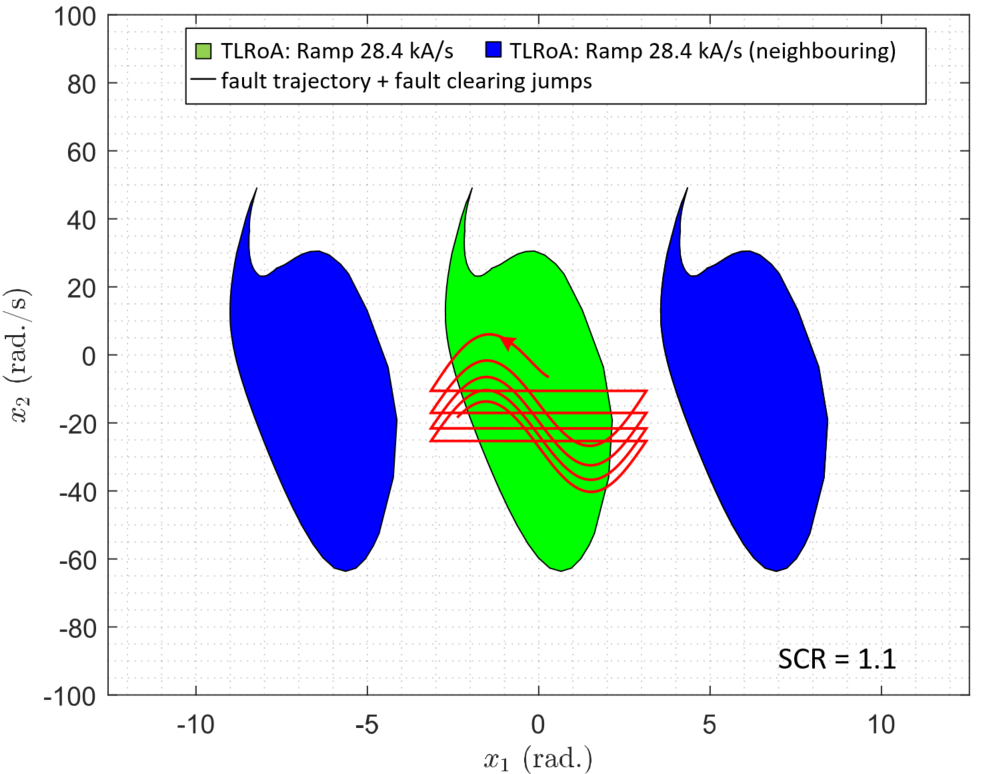}
    \caption{Transient stability assessment: Less stable system with grid SCR of 1.1.}
    \label{fig:Ramp1}
\end{figure}

\subsection{Nonlinear component modelling}
The PLL structure in \cite{14} does not include hard saturation limits. In reality, saturation blocks are employed for the integrator block at the calculated PLL frequency signal. For instance, the frequency signal may be limited to a maximum of 55Hz and a minimum of 45Hz.

To obtain the TLRoA for system (1) with PLL frequency saturation, it is crucial to establish that the system is time reversible. Fig. 13 depicts the system's forward and reverse time trajectories when the PLL frequency hits the saturation limits, i.e. $\pm5$Hz. Fig. 13 shows a clear violation of time reversal, and the deviation in the reverse trajectory is seen at the instant of frequency saturation. This asymmetry can be attributed to the violation of the prerequisites, where it was suggested that the system must be locally Lipschitz; that is, it is a continuous function, and its derivative with respect to the state variables is bounded, i.e. a saturation makes the derivate to be undefined.

\begin{figure}[h]
    \centering
    \includegraphics[width=7.5cm]{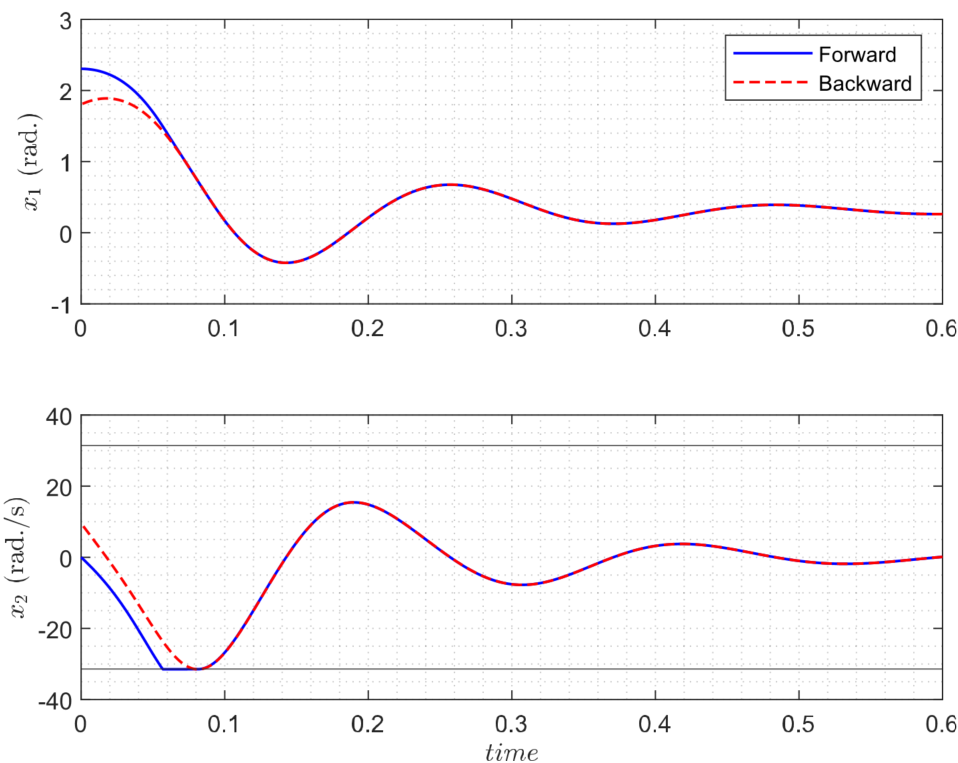}
    \caption{Violation of time reversal at the discontinuity.}
    \label{fig:Ramp1}
\end{figure}

\subsubsection{Smooth saturation characteristics}
One approach to ensure that system (1), with hard PLL frequency saturation, satisfies the established reverse time requirements is to smooth the hard saturation characteristics. To achieve this, a smooth saturation function is recommended. Our analysis evaluated several smooth functions, such as arctan, tanh, sigmoid, and logistic functions. Due to its simplicity in application, the tanh function was chosen, which has a range of (-1, 1) and can be easily scaled based on the PLL frequency limits, as follows:
\begin{equation}
     x_{2}^{sat} = x_{2}^{max} \cdot \text{tanh}( \frac{x_{2}}{x_{2}^{max}})\\
\end{equation}

Fig. 14 shows the forward and reverse trajectories of system (1) with smooth saturation (4). It is noticeable that there is a match in the forward and reverse trajectory, indicating that the system with smooth saturation satisfies the established reverse time theory requirements.

\begin{figure}[h]
    \centering
    \includegraphics[width=7.5cm]{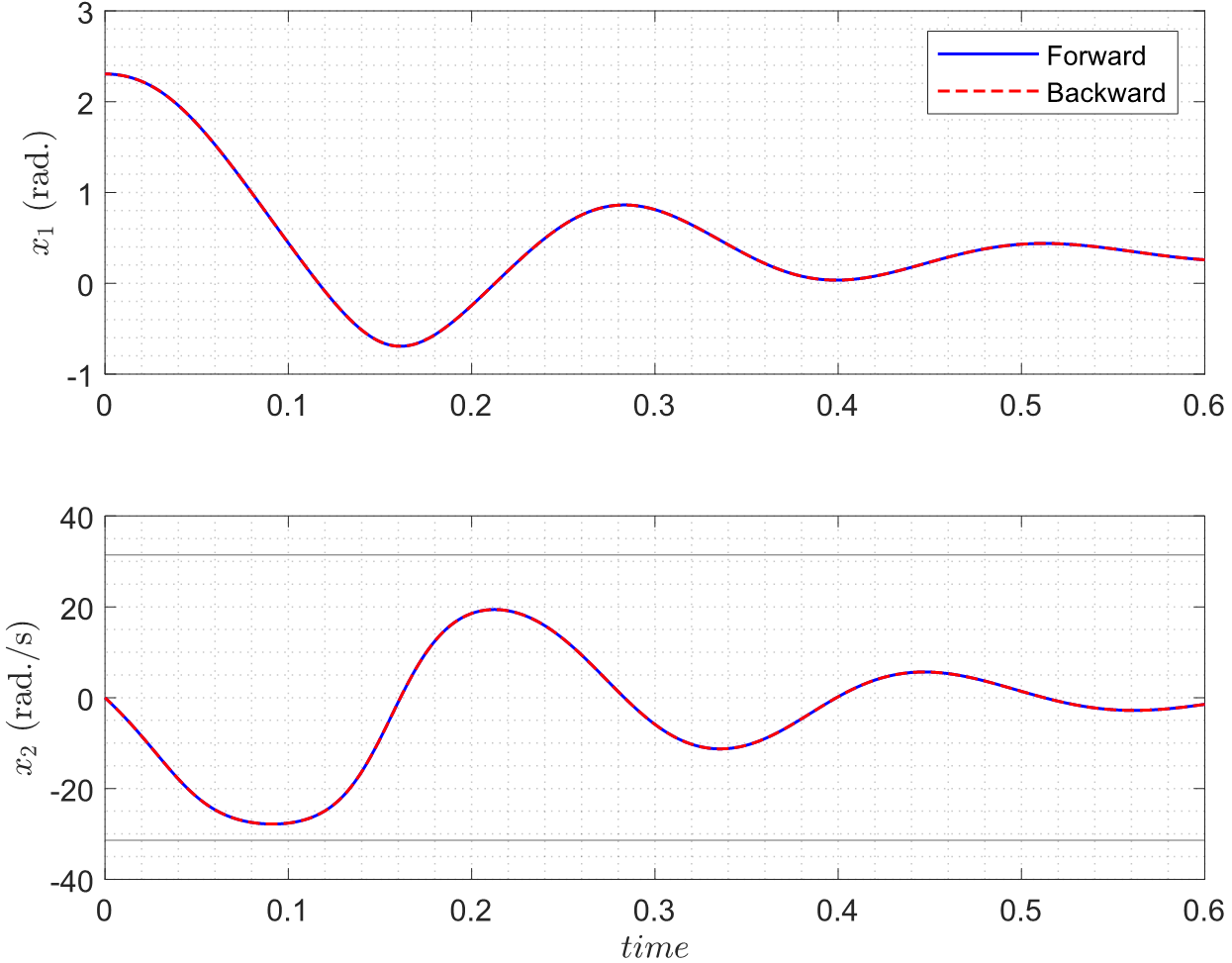}
    \caption{Characteristics of hard saturation vs smooth saturation: (a) Visualisation of smooth saturation, (b) Forward and reverse time trajectory with smooth saturation.}
    \label{fig:Ramp1}
\end{figure}

\subsubsection{TLRoA with smooth saturation}
Once it is established that the system is time reversible, the estimation of its TLRoA can be carried out. For this purpose, it is necessary to establish the initial conditions. The initial RoA is not affected by the PLL saturation since it is very small. However, during the reverse time trajectory, saturation must be considered. Therefore, considering equation (4), system (1) is transformed into a differential algebraic equation. Fig. 15 illustrates the estimated TLRoA for system (1) with smooth saturation.

\begin{figure}[h]
    \centering
    \includegraphics[width=7.5cm]{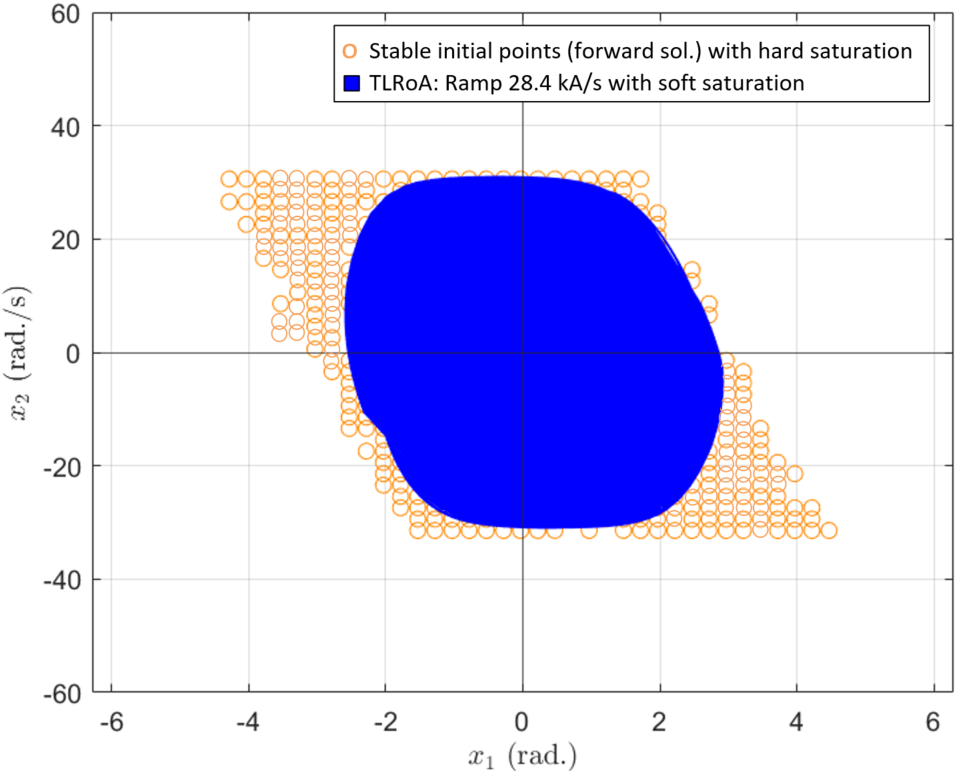}
    \caption{Estimated TLRoA for system (1) with smooth saturation.}
    \label{fig:Ramp1}
\end{figure}

Fig. 15 shows that the estimated TLRoA with smooth saturation is a subset of the forward simulated RoA with hard saturation, indicating that the proposed methodology has achieved its objective. Additionally, Fig. 16 compares the TLRoAs of the system with and without frequency saturation. One might be tempted to conclude that the system with saturation is less stable because its RoA appears smaller than the system without saturation. However, the addition of saturation ensures that the system trajectories cannot leave the RoA without saturation, making the new system more stable overall. It is important to note that the RoA without saturation is theoretical. In reality, frequency saturation is often necessary to prevent the system from becoming unstable.

\begin{figure}[h]
    \centering
    \includegraphics[width=7.5cm]{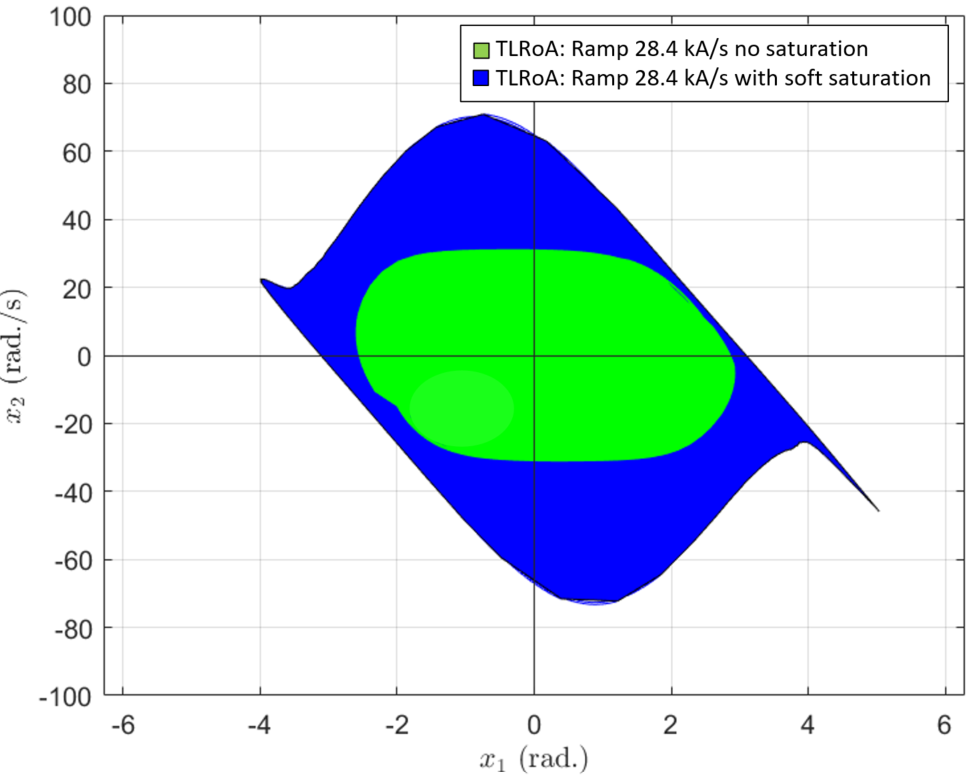}
    \caption{Estimated TLRoA for system (1) with and without saturation.}
    \label{fig:Ramp1}
\end{figure}

\subsection{Sampling for TLRoA estimation}
In Fig. 4, a red asterisk on the TLRoA boundary indicates one time-domain simulation. To estimate a smooth boundary, approximately 300 simulations were conducted. Although this number is much lower than the 3200 simulations required for the forward simulated RoA, upon analysing the distribution of sample points (red asterisks), it is possible that a reduction in the number of simulations could be achieved.

In traditional sampling, data is collected at a fixed rate, which can result in inaccurate or inefficient results if the data is not sufficiently distributed. However, adaptive sampling solves this problem by dynamically adjusting the sampling rate based on the data characteristics. For example, suppose certain areas of the TLRoA boundary are not sufficiently represented. In that case, adaptive sampling can increase the sampling in those areas to obtain a more representative boundary and vice versa.

A TLRoA estimation framework using adaptive sampling was created in a Python environment with the help of the "Adaptive" open-source Python library \cite{24}, designed for parallel active learning of mathematical functions. 

To estimate the TLRoA, multiple reverse-time simulations with a known set of initial conditions (initial RoA) were performed. Adaptive sampling aims to identify a subset of these initial conditions to estimate a smooth TLRoA boundary. 

In Fig. 17, the TLRoA estimated with different loss functions is presented. When a loss goal of 0.03 is set, the homogeneous loss function with N=50 samples performs poorly in estimating a good TLRoA because it does not consider the uneven distribution of the RoA boundary. In contrast, the curvature loss function with N=68 samples performs the best as it samples more in the curvature, leading to fewer samples and shorter computation time. The Euclidean loss function with N=183 samples took approximately 21 s to complete. Another case study was conducted to evaluate the feasibility of the homogeneous loss function with a reduced loss goal of 0.005. Despite an increased number of samples (resulting in a computation time of about 43 s), the homogeneous loss function still failed to improve the estimated TLRoA. 

\begin{figure}[h]
    \centering
    \includegraphics[width=7.50cm]{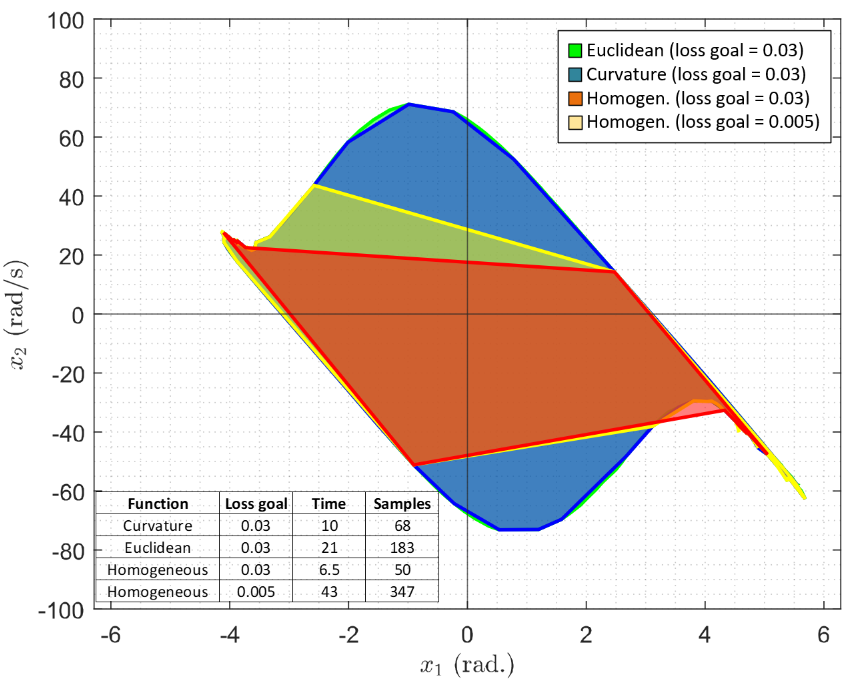}
    \caption{Adaptive sampling with various loss functions and loss goals.}
    \label{fig6}
\end{figure}

\begin{figure*}[h]
    \centering
    \includegraphics[width=15.0cm]{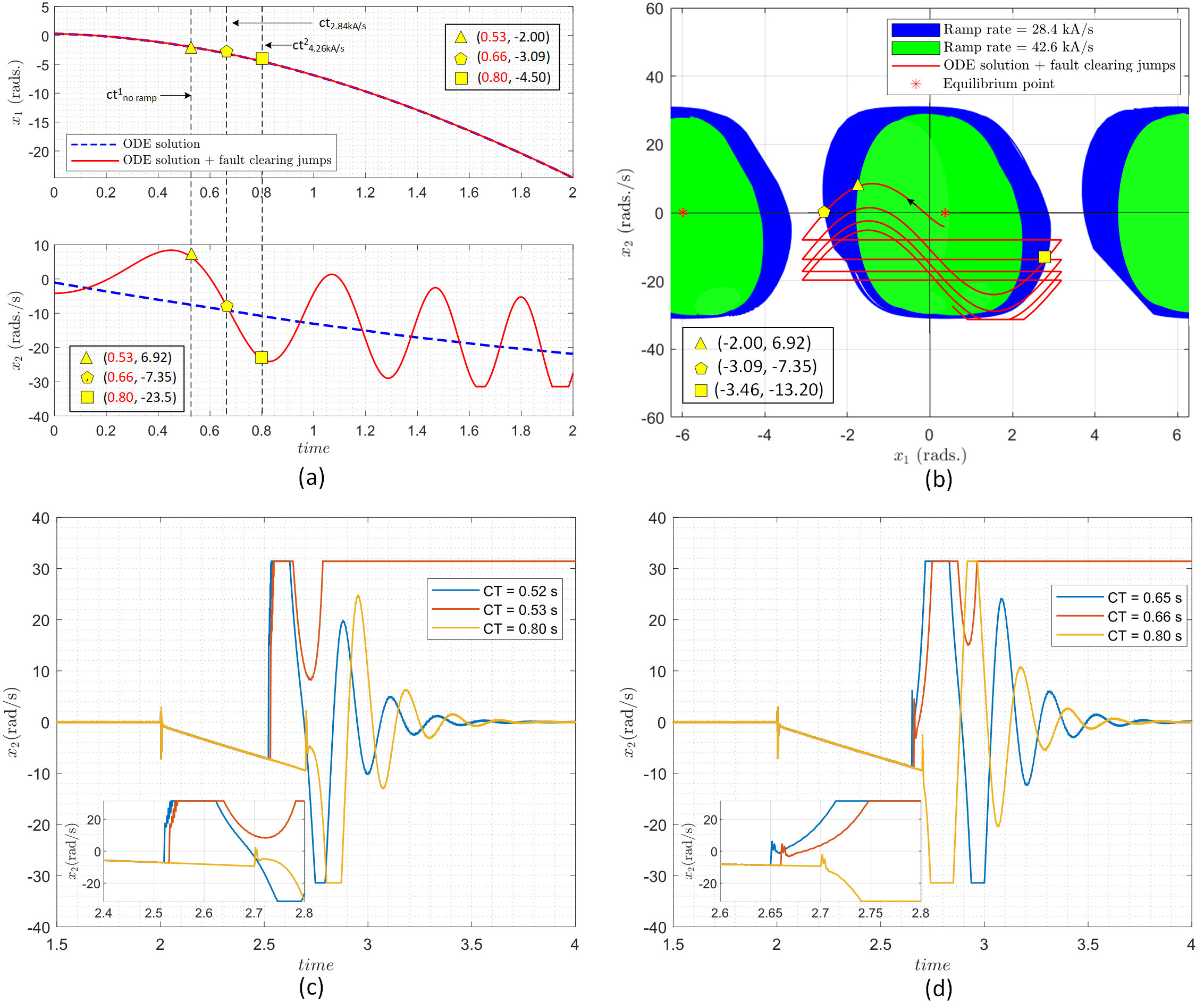}
    \caption{Time domain verification; (a) Forward fault trajectory of ODE (1) and (2), (b) Hybrid approach to assessing transient stability, (c) Fault clearing times from PSCAD simulations for a system with 28.4 kA/s, (d) Fault clearing times form PSCAD simulations for a system with 42.6 kA/s.}
    \label{fig6}
\end{figure*}

\section{Time domain verification}
In this section, we assessed our proposed methodology for transient stability through time-domain simulations employing an EMT WT switching model in PSCAD. The EMT model used the configuration specified in \cite{12}, wherein the current controller gains were tuned to achieve a fast response. The mathematical model for the WT system during and after the fault is discussed in detail in \cite{12}.

Figure 18(a) depicts a simulation (with frequency saturation) of a balanced fault, presenting a severe grid voltage dip over an extended period. The parameters used in the simulation are $V_g=0$ pu, $i_d=0.01$ pu, and $i_q=-1$ pu. 

Fig. 18(a) presents the PLL angle and frequency jumps upon fault clearance. In accordance with Section III, Fig. 18(b) illustrates the estimated TLRoA for system (1), considering two distinct post-fault active current ramp rates: 28.4 kA/s and 42.6 kA/s. Furthermore, the red curve from Fig. 18(a) is overlaid on the estimated TLRoA in Fig. 18(b) using delta-omega coordinates.

The two case studies with different post-fault active current ramp rates are carried out in PSCAD, to investigate and cross-check the theoretical critical clearing times obtained from Figs. 18(a) and 18(b). Figure 18(c) shows the time-domain simulations indicating the clearing time for the WT system with a post-fault active current ramp rate of 28.4 kA/s. The system is unstable when the fault is allowed to propagate beyond 0.52 s and cleared sometime before 0.79 s. However, the system stabilises if the fault is allowed to propagate and cleared at 0.80 s. This is in line with the theoretical understanding discussed in Figs. 18(a) and 18(b).

Similarly, Fig. 18(d) shows the time-domain simulations indicating the clearing time for the WT systems with a post-fault active current ramp rate of 42.6 kA/s. The system is unstable when the fault is allowed to propagate beyond 0.65 s and cleared sometime before 0.79 s. However, the system stabilises if the fault is allowed to propagate and cleared at 0.8 s. Again, this is consistent with the theoretical understanding.

The proposed methodology is highly reliable for investigating the transient stability of wind turbines (WTs). Compared to the traditional approach of estimating the RoA through forward-time simulation, our method has a distinct advantage. The conventional method involves guessing initial conditions, resulting in either a stable or unstable trajectory. On the other hand, our proposed methodology with adaptive sampling solves only stable trajectories (with saturation) in reverse time, leading to a faster and more efficient estimation of the RoA.  

\section{Discussion}
An extended methodology for transient stability assessment based on reverse time trajectory was introduced in this paper, and its effectiveness on a small test power system (i.e. single WT) was evaluated. The importance of establishing a solid foundation for the proposed method before extending it to larger power systems is believed to be crucial. In this regard, a thorough literature review on transient stability assessment methods was conducted, and a list of recently published papers from high-level journals addressing this topic was compiled \cite{25}-\cite{28}, with a specific focus on works that initially presented their methodologies on smaller systems before progressing to larger ones. This allows us to showcase how our research aligns with the existing body of knowledge and how we intend to build upon it. Nevertheless, the significance of exploring the behaviour of WTs when interconnected in an AC network, while considering interactions among WT converters is recognised. Subsequently, the applicability of the methodology to such scenarios will be investigated in our future research. Additionally, a more comprehensive model, including generator-side converters, beyond the simplified representation as DC voltage sources, will be provided.

\section{Conclusion}
This study builds upon previous research on nonlinear modelling and transient stability assessment of type-4 wind turbines (WTs) and introduces an extended methodology for assessing the transient stability of WT systems. The main conclusions of this paper are as follows:

\begin{enumerate}
    \item The hybrid approach based on the Lyapunov function and reverse time trajectory provides a reliable estimate of the post-fault system boundary (RoA). It was observed that the clearing times obtained from the proposed method are consistent with the results obtained from the PSCAD simulations.

    \item The sensitivity studies provided fresh insights into how system parameters influence the TLRoA. It was observed that varying the backward simulation time until the TLRoA stops growing results in a better estimate of the system boundary. Further, a slower post-fault active current ramp rate, a lower active current injection during the fault, and a higher grid SCR lead to a larger RoA.
    
    \item The proposed methodology enables modelling nonlinearities such as PLL frequency saturation while still satisfying the prerequisites of the reverse time trajectory, thereby enabling the rapid estimation of actual system RoAs with nonlinear characteristics.   
    
    \item Adaptive sampling greatly enhances the accuracy and efficiency of estimating system RoAs. To estimate a smooth boundary using forward time simulation, 3200 simulations with 30 minutes of computation time were required. However, adaptive sampling with a curvature loss function reduced the computational burden to only 68 simulations in 10 seconds.
    
\end{enumerate}

While the primary focus of this work is on individual wind turbines, our proposed method has the potential for adaptability at the wind power plant (WPP) level as well, and this is a future extension of our research.

\begin{IEEEbiography}[{\includegraphics[width=1in,height=1.25in,clip,keepaspectratio]{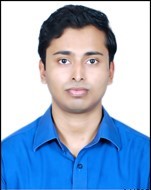}}]{Sujay Ghosh} Sujay Ghosh is currently pursuing his Ph.D. on stability of large renewable power plants in low inertia systems, from Technical University of Denmark (DTU) in collaboration with Ørsted Wind power A/S. He has a Masters in Science in Electrical engineering from DTU. His research field is in power system stability and control. Previously, he has also worked as a system studies engineer in the electrical industry in India and Denmark.
\end{IEEEbiography}

\begin{IEEEbiography}[{\includegraphics[width=1in,height=1.25in,clip,keepaspectratio]{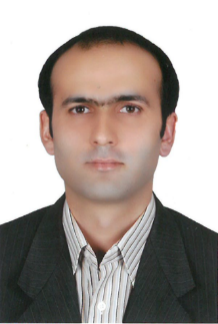}}]{Mohammad Kazem Bakhshizadeh} received the B.S. (2008) and M.S. (2011) degrees in electrical engineering from Amirkabir University of Technology, Tehran, Iran, and the PhD degree from Aalborg University, Aalborg, Denmark in 2018. In 2016, he was a visiting scholar at Imperial College London, London, U.K.
He is currently a senior power system engineer at Ørsted Wind Power, Fredericia, Denmark. His research interests include power quality, modelling, control and stability analysis of power converters, and grid converters for renewable energy systems.
\end{IEEEbiography}

\begin{IEEEbiography}[{\includegraphics[width=1in,height=1.25in,clip,keepaspectratio]{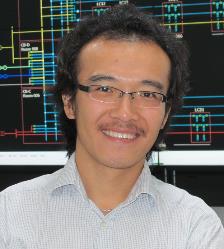}}]{Guangya Yang} (M'06--SM'14) Guangya Yang joined Technical University of Denmark (DTU) in 2009 and currently senior scientist with the Department of Wind and Energy Systems. Previously he has also been full time employee at Ørsted working on electrical design of large offshore wind farms. His research field is in stability, control, and operation of power systems. He has received numerous research grants including the recent H2020 MSCA ITN project InnoCyPES as coordinator. He is currently serving as lead editor for IEEE Access Power and Energy Society Section, and has been editorial board member of several journals. He is a member of IEC Technical committee “Wind power generation Systems” (TC88) and the convenor of IEC61400-21-5. 
\end{IEEEbiography}

\begin{IEEEbiography}[{\includegraphics[width=1in,height=1.25in,clip,keepaspectratio]{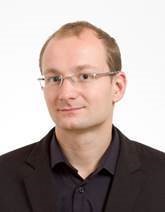}}]{Lukasz Kocewiak} (M’12–SM’16) Łukasz Hubert Kocewiak  received the BSc and MSc degrees in electrical engineering from Warsaw University of Technology in 2007 as well as the PhD degree from Aalborg University in 2012. Currently he is with Ørsted and is working as an R\&D manager. He is a power system specialist within the area of design of electrical infrastructure in large offshore wind power plants. The main direction of his research is related to harmonics, stability and nonlinear dynamics in power electronics and power systems especially focused on wind power generation units. He is the author/co-author of more than 100 publications. He is a member of various working groups / activities within Cigré, IEEE, IEC.
\end{IEEEbiography}

\vfill


\begin{thebibliography}{1}
\bibliographystyle{IEEEtran}

\bibitem{1}
GWEC, Global wind report 2022.

\bibitem{2}
D. Wang, J. L. Rueda Torres, A. Perilla, E. Rakhshani, P. Palensky and M. A. A. M. van der Meijden, "Enhancement of Transient Stability in Power Systems with High Penetration Level of Wind Power Plants," 2019 IEEE Milan PowerTech, 2019, pp. 1-6.

\bibitem{3}
Y. Li et al., "PLL Synchronization Stability Analysis of MMC-Connected Wind Farms Under High-Impedance AC Faults," in IEEE Transactions on Power Systems, vol. 36, no. 3, pp. 2251-2261.

\bibitem{4}
L. N. Arruda, S. M. Silva and B. J. C. Filho, "PLL structures for utility connected systems," Conference Record of the 2001 IEEE Industry Applications Conference. 36th IAS Annual Meeting (Cat. No.01CH37248), 2001, pp. 2655-2660 vol.4.

\bibitem{5}
J. Hu, Q. Hu, B. Wang, H. Tang and Y. Chi, "Small Signal Instability of PLL-Synchronized Type-4 Wind Turbines Connected to High-Impedance AC Grid During LVRT," in IEEE Transactions on Energy Conversion, vol. 31, no. 4, pp. 1676-1687, Dec. 2016.

\bibitem{6}
Wu, F., Zhang, X.-P., Godfrey, K., Ju, P. "Small signal stability analysis and optimal control of a wind turbine with doubly fed induction generator", (2007) IET Generation, Transmission and Distribution, 1 (5), pp. 751-760.

\bibitem{7}
Liu, H., Sun, J. "Voltage stability and control of offshore wind farms with AC collection and HVDC transmission", (2014) IEEE Journal of Emerging and Selected Topics in Power Electronics, 2 (4), art. no. 6914527, pp. 1181-1189.

\bibitem{8}
Amin, M., Molinas, M. "Understanding the Origin of Oscillatory Phenomena Observed between Wind Farms and HVdc Systems", (2017) IEEE Journal of Emerging and Selected Topics in Power Electronics, 5 (1), art. no. 7605516, pp. 378-392.

\bibitem{9}
Vu, T.L., Turitsyn, K. "Lyapunov Functions Family Approach to Transient Stability Assessment", (2016) IEEE Transactions on Power Systems, 31 (2), art. no. 7106572, pp. 1269-1277.

\bibitem{10}
M. Bravo, A. Garcés, O. D.Montoya, and C. R. Baier, “Nonlinear analysis for the three-phase PLL: A new look for a classical problem,” in Proc. IEEE Workshop Control Model. Power Electron., Padua, Italy, Jun. 2018, pp. 1–6.

\bibitem{11}
S. L. Brunton, B. W. Brunton, J. L. Proctor, and J. N. Kutz, “Koopman invariant subspaces and finite linear representations of nonlinear dynamical systems for control,”PLOS ONE, vol. 11, no. 2, pp. 1–19, Feb. 2016.

\bibitem{12}
M. K. Bakhshizadeh, S. Ghosh, Ł. Kocewiak, and G. Yang, “Improved Reduced-Order Model for PLL Instability Investigations,” IEEE Access, pp. 1–1, 2023, doi: 10.1109/ACCESS.2023.3294818.

\bibitem{13}
S. Ghosh, Md. K. Bakhshizadeh, Ł. Kocewiak and G. Yang, “Nonlinear Stability Assessment Of Type-4 Wind Turbines During Unbalanced Grid Faults Based On Reduced-Order Model”, arXiv preprint arXiv: 2306.05881, 2023. 

\bibitem{14}
M. K. Bakhshizadeh, S. Ghosh, G. Yang and Ł. Kocewiak, "Transient Stability Analysis of Grid-Connected Converters in Wind Turbine Systems Based on Linear Lyapunov Function and Reverse-Time Trajectory," in Journal of Modern Power Systems and Clean Energy, doi: 10.35833/MPCE.2023.000190.

\bibitem{15}
Lamb, J.; Roberts, J., "Time-reversal symmetry in dynamical systems: A survey", Phys. Nonlinear Phenom 1998, 112, 1–39, doi:10.1016/S0167-2789(97)00199-1.

\bibitem{16}
Ł. Kocewiak, et. al., “Overview, Status and Outline of Stability Analysis in Converter‐based Power Systems,” \emph{The 19th International Workshop on Large-Scale Integration of Wind Power into Power Systems}, Energynautics GmbH, 11-12 November 2020.

\bibitem{17}
P. Varaiya, F. F. Wu, and R. Chen, “Direct methods for transient stability analysis of power systems: Recent results,” Proc. IEEE, vol. 73, no. 12, pp. 1703–1715, Dec. 1985

\bibitem{18}
Contessa G., "Scientific models and fictional objects",
(2010) Synthese, 172 (2), pp. 215 - 229, doi: 10.1007/s11229-009-9503-2

\bibitem{19}
Nelson R.A., Olsson M.G., "The pendulum—Rich physics from a simple system",v(1986) American Journal of Physics, 54 (2), pp. 112 - 121, doi: 10.1119/1.14703


\bibitem{21}
Birkhoff G.D., "The restricted problem of three bodies",
(1915) Rendiconti del Circolo Matematico di Palermo, 39 (1), pp. 265 - 334, doi: 10.1007/BF03015982

\bibitem{22}
S. Ghosh, M. K. Bakhshizadeh, Ł. Kocewiak, and G. Yang, “Nonlinear Stability Investigation Of Type-4 Wind Turbines With Non-autonomous Behavior Based On Transient Damping Characteristics,” IEEE Access, pp. 1–1, 2023, doi: 10.1109/ACCESS.2023.3297486


\bibitem{24}
Nijholt, Bas, Weston, Joseph, Hoofwijk, Jorn, \& Akhmerov, Anton. (2023). python-adaptive/adaptive: version 1.0.0 (v1.0.0). Zenodo. https://doi.org/10.5281/zenodo.7938435

\bibitem{25}
Zhang, Y., Zhang, C., \& Cai, X., "Large-Signal Grid-Synchronization Stability Analysis of PLL-Based VSCs Using Lyapunov's Direct Method", (2020) IEEE Transactions on Power Systems, 37 (1), 788-791. https://doi.org/10.1109/TPWRS.2021.3089025

\bibitem{26}
Zhang C, Molinas M, Li Z and Cai X, "Synchronizing Stability Analysisand Region of Attraction Estimation ofGrid-Feeding VSCs UsingSum-of-Squares Programming", (2020) Front. Energy Res. 8:56. https://doi.org/10.3389/fenrg.2020.00056 

\bibitem{27}
Chen Zhang, Xu Cai, Atle Rygg, Marta Molinas, "Modeling and analysis of grid-synchronizing stability of a Type-IV wind turbine under grid faults", International Journal of Electrical Power \& Energy Systems, Volume 117, 2020, 105544, ISSN 0142-0615, https://doi.org/10.1016/j.ijepes.2019.105544.

\bibitem{28}
H. Wu and X. Wang, "Design-Oriented Transient Stability Analysis of PLL-Synchronized Voltage-Source Converters", in IEEE Transactions on Power Electronics, vol. 35, no. 4, pp. 3573-3589, April 2020.

\end{thebibliography}
\end{document}